# Cold Quark-Gluon Plasma EOS Applied to a Magnetically Deformed Quark Star with an Anomalous Magnetic Moment

Keith Andrew[1], Eric Steinfelds[1,2], Kristopher Andrew[3]
[1]Department of Physics and Astronomy, WKU
Bowling Green, KY 42101
[2]Department of Physical Sciences, Alverno College,
Milwaukee, WI 53215
[3]Department of Science, Schlarman Academy
Danville, IL 61832

**Abstract**

We consider a QCD cold plasma motivated Equation of State (EOS) to examine the impact of an Anomalous Magnetic Moment (AMM) coupling and small shape deformations for static oblate and prolate core shapes of quark stars. Using the Fogaça QCD motivated EOS which shifts from the high temperature low chemical potential quark gluon plasma environment to the low temperature high chemical potential quark stellar core environment we consider the impact of an AMM coupling with a metric induced shape deformation parameter in the TOV equations. The EOS is developed using a hard gluon and soft gluon decomposition of the gluon field tensor using a mean field effective mass for the gluons. The AMM is considered using the Dirac spin tensor coupled to the EM field tensor with quark flavor based magnetic moments. The shape parameter is introduced in a metric ansatz that represents oblate and prolate static stellar cores for modified TOV equations. These equations are numerically solved for the final mass and radius states representing the core collapse of a massive star with a phase transition leading to an unbound quark-gluon plasma. We find that the combined shape parameter and AMM effects can alter the coupled EOS-TOV equations resulting in an increase in the final mass and a decrease in the final equatorial radius without collapsing the core into a black hole and without violating causality constraints, we find maximum mass values in the range: $2.3\, M_\odot < M < 2.7\, M_\odot$. These states are consistent with some astrophysical high mass magnetar/pulsar and gravity wave systems which may provide evidence for a core that has undergone a quark-gluon phase transition such as PSR 0943+10 and the secondary from the GW 190814 event.

## I. Introduction

There has been great interest in finding more realistic QCD based Equations of State (EOS) for compact objects with the discovery of anomalous compact astrophysical objects, such as the stellar remnant from the energetic hypernova 2006gy[1]. This event, caused by the sudden collapse of a high mass star, gave rise to final state mass and radius values that do not agree very well with standard compact stellar core models. The ongoing observational datasets also include the anomalous X-ray pulsar[2] RX J1856[3,4] and pulsars such as PSR 0943+10[5], PSR B1828-11[6], PSR Jo751+1807[7], PSR B1642-03, PSR J1614-2230 and PSR J0348-0432 along with potential magnetar candidates[8] where masses, radii, dual luminosity peaks and cooling rates do not always give excellent fits to conventional models based upon degenerate neutron matter. As the observational data has become more robust and refined more detailed models have emerged to help understand the varied mechanisms at play in dense QCD matter. Several models are gaining support from the observations of the Quark Gluon Plasma, QGP, demonstrating the existence of a high temperature, low chemical potential state of unconfined quarks as seen at SPS[9], LHC[10] and RHIC[11,12] laboratories. In recent years the QGP data has expanded to include the determination of relativistic QCD based transport coefficients[13] and measurements of the overall bulk viscosity for free quark matter.[14,15] Observations of X-ray, γ-ray and quasi periodic objects such as SAS J1808.4 -3658, which has a 2.49 ms period,[16] have been used to understand how the environment[17] and ambient mass density impacts fundamental QCD parameters and potentials[18]. The Ligo-Virgo detectors are also observing events that appear to correspond to massive compact secondaries such as the GW 190814[19] event indicating a secondary with a mass near or above 2.5 $M_\odot$. However, it is difficult to find a unique event without perhaps reasonable competing scenarios. For instance, in the GW 190814 event scenarios can involve the 23 $M_\odot$ primary with a secondary neutron star that is rapidly rotating[20] and could form a black hole by itself or there may be a tertiary object[21] involved which then gives mass values near the conventional neutron star limit of 2.3 $M_\odot$ for the secondary.

Various EOS have been used linking the strong coupling constant and quark masses to the local density function using the real time formalism and the Dyson-Schwinger gap equation at finite temperature[22]. In this way high density compact cores can be used to give the density dependent strong coupling that can be expressed as an analytic function that is suitable for EOS and phase transition analysis[23] with high chemical potentials and lower temperatures. For such a case the overall charge, density, and chemical potentials are constrained by overall charge neutrality for particle number density $n_j$, mass density $\rho_f$ and electric charge $q_j$ for particle type $j$ or flavor $f$ where: $\sum q_f n_f = q_e n_e$ and $\rho_B = \rho_u + \rho_d + \rho_s$. An additional constraint from the three light quark flavor weak interaction equilibrium condition from quark interactions $d \leftrightarrow u + e^- + \bar{\nu}_e$, $s \leftrightarrow u + e^- + \bar{\nu}_e$, $s + u \leftrightarrow d + u$ constrain the chemical potentials where the neutrinos and antineutrinos exit the core on a short time scale compared to the core formation process effectively causing their chemical potentials to vanish resulting in $\mu^d = \mu^u + \mu^e$, $\mu^d = \mu^s$. Then the EOS can be expressed as a pressure function that is related to the chemical potential from $P = -\partial U/\partial V = n^2 \partial(\varepsilon/n)/\partial n = n\mu - \varepsilon$ where $\varepsilon$ is the energy density and the chemical potential is $\mu = \partial \varepsilon / \partial n$. For the massless quark case this gives a direct comparison to the MIT bag model with bag constant $\mathcal{B}$ where $3P = (\varepsilon - 4\mathcal{B}) = \varepsilon - 4(3\mu^4/4\pi)$ linking [24] the vacuum pressure bag constant to the quark chemical potentials which are energy dependent. Other compact core models have focused on modified EOS for various states of dense matter in the QCD ground state of Color Flavor Locked (CFL) dense matter[25,26] with 2-d and 3-d color superconductivity[27,28,29] and in the Nambu-Jona-Lasinio (NJL) model[30]. Many models admit to a comparison to versions of the generalized MIT Bag model[31] approach and include QCD mean field approximations[32,33] along with computationally intensive numerical lattice approaches that are being improved to accommodate finite temperature chemical potentials.[34] Refined models include crustal effects[35], deformations[36], layering[37], cooling rates[38], strange quark cores[39], post-merger analysis[40], finite temperature boson stars[41], kaon condensation,[42,43] superconducting color vector potential effects[44,45], phase transitions, and different crystalline cores[46,47,48].

In addition, collapsed stellar cores are in general rotating and often possess very strong magnetic fields[49]. For collapsed stellar cores composed of self-bound quark matter the magnetic field energy density should not exceed the vacuum bound state energy density for a quark system with an energy of order 1GeV within a diameter of about 1fm which results in a maximum magnetic field on the order of:

$$\frac{B_{max}^2}{2\mu_o} \sim \varepsilon_{binding} \rightarrow B_{max} \sim \sqrt{2\mu_o \varepsilon_{binding}} \sim 10^{15} T. \quad (1)$$

Detailed mechanisms for achieving magnetic fields of this magnitude are not well understood and high resolution data from the NASA Neutron Star Interior Composition Explorer (NICER) has given a surface map of PSR J0030+0451 showing a complex multipole field structure.[50] A simple and direct method for estimating the magnetic field strength of a collapsed star is to assume that the collapse is magnetic flux conserving from the initial state to the final state, where strong main sequence progenitor magnetic fields can be ~ 1T, with a radius of 1 Gm, at 10 M$_\odot$ and collapse to a radius of order 10 km giving:

$$\phi_{B_i} = \oint \vec{B} \cdot d\vec{A}_i = BA_i = BA_f = \phi_{B_f} \rightarrow B_f = \left(\frac{r_i}{r_f}\right)^2 B_i \sim 10^{14} T. \quad (2)$$

However, these overly optimistic estimates do not account for significant mass loss that would make this estimate untenable for most cases. If we consider the size cutoff to be the Schwarzschild radius, $R=2GM/c^2$, when the magnetic field energy density and the gravitational field energy density are the same then we find similar maximum field strengths:

$$\varepsilon_B = \frac{B_{max}^2}{2\mu_o} \sim \frac{9GM^2}{20\pi R^4} = \varepsilon_{Gravitational} \rightarrow B_{max} \sim \sqrt{\frac{9\mu_o c^8}{160\pi G^3 M^2}} \sim 10^{15} T. \quad (3)$$

Here we examine the low temperature interactions that can be expressed as hard short wavelength gluons, and soft long wavelength gluons, with an EOS that includes a magnetic field and where the static compact core is distorted into an oblate or prolate shape necessitating the use of the modified TOV equations. One expects the quarks and gluons to contribute differently to the EOS due to their different spin and color degrees of freedom. For a compact core at the Fermi temperature, estimated by assuming a quark separation on the order of a fraction of a neutron with number density $n$, and quark mass m, then the Fermi temperature is $T \sim \hbar^2 n^{2/3} / 3mk \sim 150 MeV$ where $\hbar$ is the reduced Planck's constant and k is the Boltzmann constant. If we consider the pressure for noninteracting particles expressed as being due to free spin ½ fermions and the pressure due to massless spin 1 bosons for an SU(3)$_c$ color symmetry where the number of degrees of freedom depends on spin, color and flavor degrees of freedom then the pressures are

$$p_{gluons} = \frac{g_{gluons}\pi^2}{90}T^4 \quad g_{gluon} = 2_{spin} \cdot 8_{color} = 16 \quad p_{quarks} = \frac{g_{quarks}7\pi^2}{360}T^4 \quad g_{quark} = 2_{spin} \cdot 3_{color} \cdot 2_{flavor} = 12 \quad \frac{p_{gluons}}{p_{quarks}} \sim 4 \quad (4)$$

indicating that the gluons can play a considerable role in the total pressure of the compact core. The magnetic field resulting from the internal current densities also contributes to the pressure via a Lorentz force density which, by applying Maxwell's equations, can be expressed as $F_j = \varepsilon_{jkl} j_k B_l = -\partial_i T_{ji}$ where $T_{ji} = -(1/2\mu_o) [B^2 \delta_{ji} - 2B_j B_i]$ resulting in pressure like forces both parallel and perpendicular to the magnetic field direction giving rise to anisotropic magnetic field interactions[51]. For the observed field strengths these magnetic pressure terms can be similar in magnitude to the quark and gluon pressure terms at the Fermi temperature of the stellar core: $p_{mag}/p_{quark} \sim 0.5$. Fields of this strength will alter the EOS and can allow for a more massive core where crustal rigidity and rapid rotation[52] could deform[53] the stellar core into a prolate or oblate shape and can alter the long-term stability of the star[54]. For instance, Tatsumi has found that when all the quarks stay in the lowest Landau level there is a critical magnetic field where the EOS can support more mass in a static spherical geometry.[55] However in general the magnetic field will alter the core shape to be oblate or prolate, which, for neutron stars, would result in small changes in the equatorial radius and maximum mass. For neutron and quark stars a static shape deformation can be explored with modified TOV equations by introducing a shape parameter in the metric[56].

To explore the resulting changes in the EOS with a magnetic field we look at separating the QCD interaction into soft and hard gluon modes as is often used for a cold quark-gluon QCD condensate[57,58] method developed by Fogaça[59] with magnetic field terms that exhibit an external field and an internal field coupling between the anomalous magnetic moment, AMM, of the quarks and the magnetic field of the form $\sigma^{\mu\nu} F_{\mu\nu}$ where $\sigma_{\mu\nu}$ is the spin tensor that is proportional to the commutator of the Dirac matrices and $F_{\mu\nu}$ is the EM field tensor in a fashion similar to Mao[60]. Supporting the more massive and smaller compact core consistent with observations that can be formed by the deconfinement of the quarks requires a modified EOS. Here we examine such an EOS that results in changes in the maximum mass and radius values for quark stars by including both the magnetic couplings in the EOS and the shape parameter generalized TOV equations. In the next section we summarize the QCD motivated quark gluon condensation EOS with a magnetic field and AMM coupling applied in the mean field followed by the TOV equations. We then numerically solve for the mass-radius dependence and discuss the resulting values.

## II. QCD Motivated EOS

The EOS of interest here is the cold QGP model of Fogaça modified to include the AMM quark interactions and magnetic field. The method provides a way to extend the high temperature low chemical potential system to a low temperature high chemical potential stellar core by decomposing the gluon field into high momentum hard gluons and low momentum soft gluons. The hard gluon terms are simplified in the mean field limit and the soft gluons are replaced by matrix elements of the plasma gluon field. Here we consider our system to be a remnant quark stellar core that is in weak force equilibrium and in an electric charge neutral state. To find the gluon and quark pressure terms consider the QCD-Core Lagrangian for interacting quarks with flavors denoted by f, and the SU(3) generators given by $T^a$ with structure constants $f^{abc}$, where a is the color index q is the EM charge, g is the strength of the color coupling, and $\kappa_f = \text{diag}(a_u, a_d)$ is the AMM coupling, given as

$$\mathcal{L}_{core} = -\frac{1}{4} H^a_{\mu\nu} H^{a\mu\nu} - \sum_{f=1}^{N_f} \bar{\psi}^f_i \left[ i\gamma^\mu \left( \delta_{ij} \partial_\mu - ig T^a_{ij} G^a_\mu \right) - \delta_{ij} m_f + \kappa_f \delta_{ij} F_{\mu\nu} \sigma^{\mu\nu} \right] \psi^f_j \quad (5)$$

where the EM tensor, the gluon field tensor, and the spin tensors are defined as

$$F_{\mu\nu} = \partial_\mu A_\nu - \partial_\nu A_\mu$$
$$H^{a\mu\nu} = \partial^\mu G^{a\nu} - \partial^\nu G^{a\mu} + g f^{abc} G^{b\mu} G^{c\nu}$$
$$\sigma^{\mu\nu} = \frac{i}{2}[\gamma^\mu, \gamma^\nu] \qquad (6)$$
$$\kappa_f = \alpha_f \mu_B, \quad \alpha_f = \frac{\alpha_e q_f^2}{2\pi}, \quad \mu_B = \frac{e}{2m_f}, \quad \alpha_e = \frac{e^2}{4\pi}$$

for Dirac matrices in the standard representation with the magnetic field aligned in the positive z direction using electromagnetic 4-vector potential component in the y-direction $A_\mu = (A_0, A_1, A_2, A_3) = (0, 0, xB, 0)$. Initially we consider a system with one flavor and all quarks with the same mass and will later add the contributions for each flavor. We decompose the gluon field into low momentum, $A^{a\mu}$, or soft gluons, and high momentum, $\alpha^{a\mu}$, or hard gluons, using the binomial ansatz

$$G^{a\mu} = A^{a\mu} + \alpha^{a\mu} \qquad (7)$$

where the long wavelength low momentum gluons are responsible for the long-range strong force and the short wavelength high momentum gluons give rise to the short-range strong force. Assuming high gluon occupation numbers at the high cold plasma density and taking the hard gluon term to be a function of coordinates then, in the mean field approximation, the gluons in the core can be described by the conditions:

$$\alpha_\mu^a = \alpha_0^a \delta_{0\mu}$$
$$\partial^\mu A^{a\nu} = \partial^\mu \alpha^{a\nu} = 0 \quad . \qquad (8)$$

The mean field terms have nonzero matrix elements that arise from the quadratic field terms and the vector potential terms, and the gluons act with an effective mass $m_G$, resulting in a total pressure[61] as a function quark number density n, quark mass $m_q$, strong coupling g, quark degrees of freedom $\gamma_Q$ =(2 sf) for spin s and flavor f, quark AMM expressed in terms of the quark charge $q_f$, electron charge, e, $\alpha_e$, and magnetic field B expressed as

$$p = \left[\left(\frac{3g^2}{2m_G^2}\right)n^2 - \frac{251 m_G^4}{288 g^2}\right] + \frac{\gamma_Q}{2\pi^2}\left\{\frac{\pi^2 n}{2\gamma_Q}\sqrt{\left(\frac{2\pi^2 n}{\gamma_Q}\right)^{2/3} + m_q^2} - \left(\frac{2\pi^2 n}{\gamma_Q}\right)^{1/3}\frac{3m_q^2}{8}\sqrt{\left(\frac{2\pi^2 n}{\gamma_Q}\right)^{2/3} + m_q^2} + \frac{3m_q^4}{8}\ln\left[\left(\frac{2\pi^2 n}{\gamma_Q}\right)^{1/3} + \sqrt{\left(\frac{2\pi^2 n}{\gamma_Q}\right)^{2/3} + m_q^2}\right] - \frac{3m_q^4}{16}\ln(m_q^2) + \frac{\alpha_e q_f^2 e}{4\pi m_q}B\right\} + \frac{B^2}{8\pi}$$

$$p = (p_\alpha + p_A)_{Gluon} + (p_Q + p_{AMM}) + p_B = P_{Gluon-Total} + \sum_f P_{f-Quark-Total} + P_B \quad .$$
$$(9)$$

which corresponds to a sum of the hard gluon term, the soft gluon term, the quark term, which includes the AMM term, and the magnetic field term, where the total quark pressure can be found by summing over each flavor for u, d, and s quarks – quarks heavier than the s quark give rise to radial oscillation instabilities and are not included in the sum[62]. Here the baryon number density is $n_B=(n_u+n_d+n_s)/3$ and nuclear number density is $n_0=0.17$ fm$^{-3}$. For numerical studies we consider the optimized values of Fogaça: $m_q=0.03$ GeV and for separate quark terms we take $m_u=4$ MeV, $m_d=7$ MeV and $m_s=150$ MeV, the gluon effective mass is $m_G = 0.4$ GeV, and $\gamma_Q=3$, for spin ½ and flavor, f = 3 degrees of freedom for u, d, s quarks and the strong coupling in this regime is given by $\alpha_s = g^2/4\pi = 0.01$ or $g = 2.7$. To characterize the overall stiffness and causality of the EOS we also calculate the speed of sound in the core and the compressibility k given by

$$c_s^2 = \frac{\partial P}{\partial \varepsilon} = \frac{\partial P}{\partial \rho}\frac{\partial \rho}{\partial \varepsilon} = k\frac{\partial \rho}{\partial \varepsilon} \quad \text{where} \quad k = \frac{\partial P}{\partial \rho}. \quad (10)$$

For comparison MIT Bag Model corresponding to a free quark no gluon core with the bag constant denoted by $\mathcal{B}$, the pressure, baryon density, speed of sound and compressibility are given as

$$p_0 = \frac{1}{3}\left(\frac{3}{2}\right)^{7/3}\pi^{2/3}\rho_B^{4/3} - \mathcal{B}, \quad \rho_B = \frac{1}{3}(\rho_u + \rho_d + \rho_s), \quad c_s = \sqrt{\frac{1}{3}}, \quad \kappa = \frac{1}{9}\left(\frac{3}{2}\right)^{7/3}\pi^{2/3}\rho_B^{1/3}. \quad (11)$$

When the gluon terms and quark masses are set to zero then Eq. (9) and Eq. (11) agree by identifying the baryon density as one third of the quark density with no bag constant, this is the classic polytrope often used for high mass white dwarf stars with degenerate relativistic matter with $p \sim \rho^{4/3}$. With soft gluons and massless quarks, the gluons produce a term that corresponds to the bag constant given by the effective gluon mass. Instead of using a mean quark mass Eq. (9) can be extended by including a quark term for each flavor using the effective mass of each quark. Using these values in Eq. (9) yields an EOS that can be used with the TOV equations to find end state mass and radius values for the three-flavor cold quark plasma stellar core with an AMM quark term. In Fig. (1) we plot EOS pressures and comparison values of dimensionless compressibility factors and sound speeds for a range of typical number densities expected in the core.

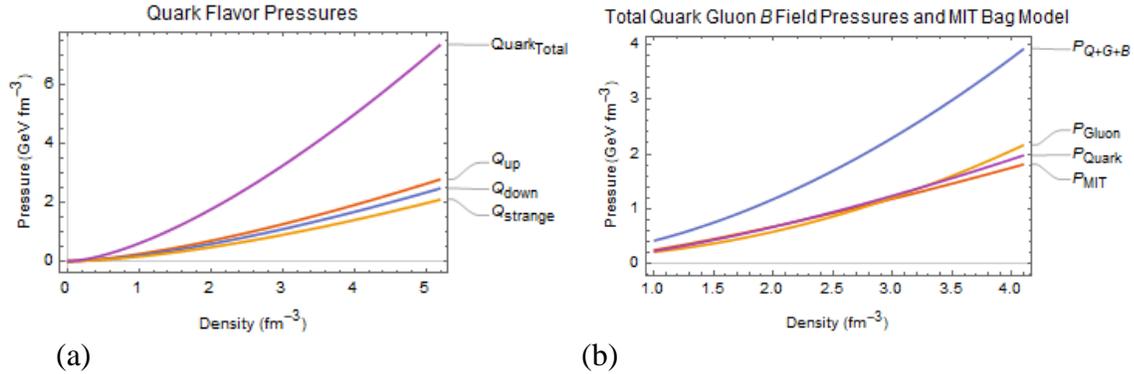

(a)          (b)

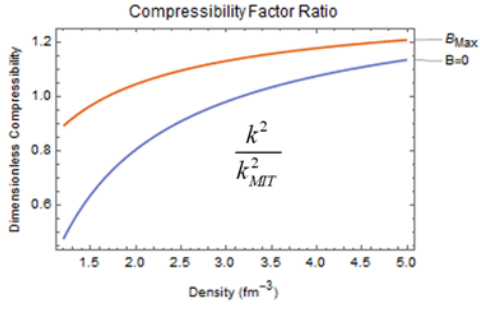
(c)

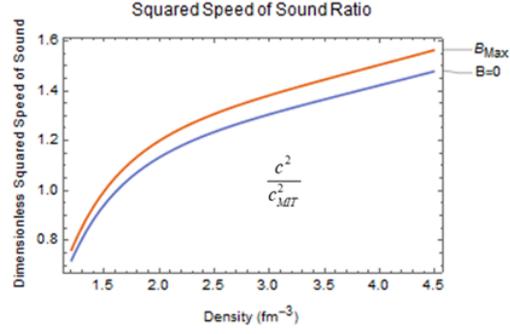
(d)

Fig. (1) (a) Each quark flavor pressure and the total core pressure showing a weak dependence on quark mass values. (b) The total quark, gluon and magnetic fields pressures shown with the individual quark, gluon, and MIT bag model pressures for a magnetic field at $10^{15}$ T and the bag model constant at 0.2 GeV/fm$^3$. (c) Compressibility factors expressed as dimensionless ratios compared to the MIT bag model with a vacuum bag constant of 0.2 GeV/fm$^3$ for no magnetic field and for the maximum field of $10^{15}$ T, (d) dimensionless squared speed compared to the MIT bag model for the case of no magnetic field and the same bag constant and for the case of the maximum magnetic field of $10^{15}$ T as a function of core density.

The numerical study of the EOS shows the importance of the gluon terms where the soft gluons establish the effective bag constant and the hard gluons display a quadratic dependence on density. Each flavor term contributes to the pressure with the lighter quarks contributing slightly more pressure and as the density increases the compressibility and the speed of sound increase.

## III. TOV Framework

We adopt the method of Zubairi[63] and Yanis[64] which includes deformation and an anisotropic magnetic field, which is similar to the methods developed by Thorne and Hartle[65] for the gravitational field for a static axisymmetric oblate/prolate ellipsoidal mass. In this case the polar z direction is scaled by a deformation parameter γ so that z = γr where r is the equatorial radial direction. Then an oblate shape corresponds to γ < 1, a prolate shape corresponds to γ > 1 and a spherical shape corresponds to γ = 1. This parametrization allows for the field equations to maintain an analytic form as modified TOV equations. The Einstein field equations can be found using the metric ansatz,

$$ds^2 = g_{\mu\nu}dx^\mu dx^\nu = -e^{2\Phi(r)}dr^2 + \left(1 - \frac{2m(r)}{r}\right)^{-\gamma} dr^2 + r^2 d\theta^2 + r^2 \sin^2\theta\, d\varphi^2 \qquad (12)$$

with Einstein and stress-energy tensors given by

$$G_{\mu\nu} = R_{\mu\nu} - \frac{1}{2}g_{\mu\nu}R = 8\pi T_{\mu\nu} \qquad (13)$$

$$T_{\mu\nu} = (\varepsilon + P)u_\mu u_\nu - g_{\mu\nu}P$$

the resulting modified TOV equations for the pressure and mass are

$$\frac{dP(r)}{dr} = -\frac{(\rho(r)+P(r))\left[\frac{r}{2}+4\pi r^3 P(r) - \frac{r}{2}\left(1-\frac{2m(r)}{r}\right)^\gamma\right]}{r^2\left(1-\frac{2m(r)}{r}\right)^\gamma} \quad (14)$$

$$\frac{dm(r)}{dr} = 4\pi r^2 \gamma \rho(r) \quad (15)$$

where the EOS, Eq.(9), give the pressure term from the effective quark-gluon Lagrangian, the radius of the object is defined to be where the pressure vanishes, and the total mass M is the product of the deformation constant and the mass inside the zero-pressure boundary, $M = \gamma\, m$.

IV. Numerical Solutions

We numerically solve the coupled system of equations: Eq. (9), Eq. (14), and Eq. (15) for the masses and radii of compact cores for the spherical, prolate and oblate cases including the impact of the magnetic field terms.

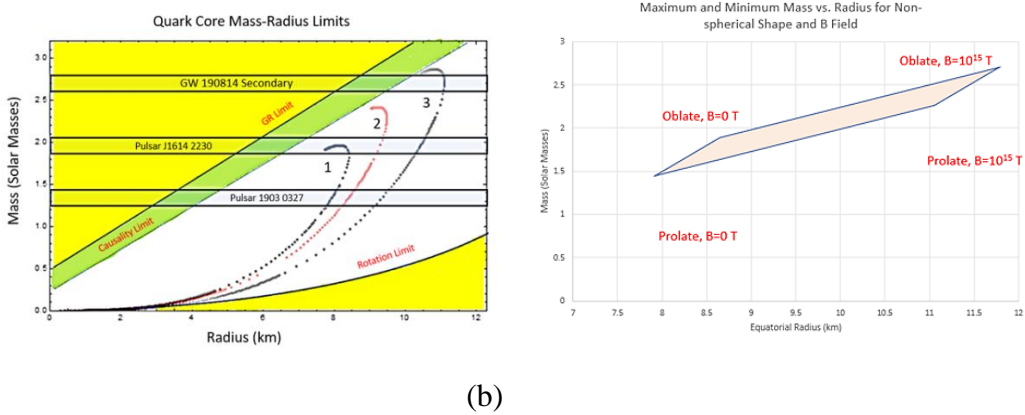

(a)                   (b)

Fig. 2 a) Quark core mass vs. equatorial radius relation from the TOV equations for an oblate deformation, $\gamma = 0.85$ and a magnetic field $B=10^{15}$ T, designated as case (1), for no deformation, $\gamma=1$, with a magnetic field $B=10^{15}$ T, designated as case (2), prolate $\gamma=1.15$ with a magnetic field, designated as case (3) and in (b) the change in the mass and equatorial radius of the core is displayed for the range of deformation values from an oblate ($\gamma=0.85$) to prolate ($\gamma=1.15$) shape and the magnetic field range from $B=0$ T to $B=10^{15}$ T.

The numerical solutions to the modified TOV equations show that the magnetic field can produce a stiffer EOS up to the critical value while the deformation increases the stiffness and mass for oblate shapes and decreases the stiffness and mass for prolate shapes.

## V.  Conclusions

We have examined the combined effects of AMM and shape deformation on a cold compact quark stellar core where the EOS is subject to a gluon field decomposition into hard and soft gluons with an additive quark term. The soft gluons produce a QCD vacuum barrier that corresponds to the bag constant, and the hard gluons produce a quadratic density dependent pressure while the quarks yield a polytropic pressure. As the deformation parameter goes from oblate to prolate the equatorial radius is larger for the same mass. As the AMM linear magnetic effect and the external field quadratic term is increased the EOS is stiffer and can support more mass for the same equatorial radius. At the highest possible magnetic field with an oblate shape the final mass and radius states are indicative of the values that have been observed in cases similar to PSR 1903 0327. As a result, high magnetic fields, gluon exchange, and oblate deformations can stiffen the EOS allowing for a larger quark star with more mass than a neutron star with densities close to ten times greater than neutron star densities. An interesting extension of this work would be to explicitly examine the effects of quark flavor chemical potentials that are expected to further soften the EOS resulting in a larger equatorial radius for a fixed mass and deformation, we are also exploring the impact of the effective massive gluon term on the cooling rate of the quark core which can be compared to data from objects such as 3C 58[66] and Cassiopeia A[67].

[61] Ibid, 54

[62] Weber, Fridolin. "Strange quark matter and compact stars." *Progress in Particle and Nuclear Physics* 54, no. 1 (2005): 193-288.

[63] Zubairi, Omair, William Spinella, Alexis Romero, Richard Mellinger, Fridolin Weber, Milva Orsaria, and Gustavo Contrera. "Non-spherical models of neutron stars." *arXiv preprint arXiv:1504.03006* (2015).

[64] Yanis, A., and Anto Sulaksono. "Deformation and anisotropic magnetic field effects on neutron star." In *AIP Conference Proceedings*, vol. 2023, no. 1, p. 020009. AIP Publishing LLC, 2018.

[65] Hartle, James B., and Kip S. Thorne. "Slowly rotating relativistic stars. II. Models for neutron stars and supermassive stars." *The Astrophysical Journal* 153 (1968): 807.

[66] Slane, Patrick O., David J. Helfand, and Stephen S. Murray. "New constraints on neutron star cooling from Chandra observations of 3C 58." *The Astrophysical Journal* 571, no. 1 (2002): L45.

[67] Ho, Wynn CG, Yue Zhao, Craig O. Heinke, D. L. Kaplan, Peter S. Shternin, and M. J. P. Wijngaarden. "X-ray bounds on cooling, composition, and magnetic field of the Cassiopeia A neutron star and young central compact objects." *Monthly Notices of the Royal Astronomical Society* 506, no. 4 (2021): 5015-5029.